\documentclass{osa-article}

\journal{osac}

\articletype{Research Article}

\usepackage{lineno}

\usepackage{float}

\usepackage{dsfont}
\usepackage{amsfonts}
\usepackage{amsmath}
\usepackage[all,poly]{xy}
\usepackage{multirow}
\usepackage{algorithm}
\usepackage{algorithmic}
\usepackage{color}
\usepackage{cite}
\usepackage{subfigure}
\usepackage{tikz,pgf}
\usepackage{array}
\usepackage[colorinlistoftodos]{todonotes}

\usetikzlibrary{fit}
\usetikzlibrary{arrows,automata}

\usepackage{verbatim}
\usetikzlibrary{positioning,shapes.geometric}
\usepackage{url}
\usepackage[T1]{fontenc}
\DeclareMathOperator*{\argmax}{argmax} %
\DeclareMathOperator*{\argmin}{argmin} %
\newcommand{\figwidth}{\columnwidth}
\usepackage{multicol}

\def\bsd{{\boldsymbol{d}}}

\def\bsg{{\boldsymbol{g}}}
\def\bsh{{\boldsymbol{h}}}

\def\bsm{{\boldsymbol{m}}}

\def\bsu{{\boldsymbol{u}}}

\def\bsw{{\boldsymbol{w}}}

\def\bsB{{\boldsymbol{B}}}

\def\bsH{{\boldsymbol{H}}}

\def\bsM{{\boldsymbol{M}}}

\def\bsR{{\boldsymbol{R}}}
\def\bsS{{\boldsymbol{S}}}

\def\bsW{{\boldsymbol{W}}}

\def\bsY{{\boldsymbol{Y}}}

\newcounter{algo}
\renewcommand{\thealgo}{\arabic{algo}}

\begin{document}

\title{3D Target Detection and Spectral Classification for Single-photon LiDAR Data }

\author{Mohamed Amir Alaa Belmekki,\authormark{1}, Jonathan Leach,\authormark{1}, Rachael Tobin,\authormark{1}, Gerald S. Buller,\authormark{1}, Stephen McLaughlin,\authormark{1}, and Abderrahim Halimi,\authormark{1,*}}


\address{\authormark{1}School of Engineering and Physical Sciences, Heriot-Watt University, Edinburgh, EH14 4AS, United Kingdom. } 

\email{\authormark{*} a.halimi@hw.ac.uk} 
 

\begin{abstract} 
3D single-photon LiDAR imaging has an important role in many applications. 
However, full deployment of this modality will require the analysis of low signal to noise ratio target returns and very high volume of data. This is particularly evident when imaging through obscurants or in high ambient background light conditions.
This paper proposes a multiscale approach for 3D surface detection from the photon timing histogram to permit a significant reduction in data volume. The resulting surfaces are background-free and can be used to infer depth and reflectivity information about the target. We demonstrate this by proposing a hierarchical Bayesian model for 3D reconstruction and spectral classification of multispectral single-photon LiDAR data. The reconstruction method promotes spatial correlation between point-cloud estimates and uses a coordinate gradient descent algorithm for parameter estimation.    
Results on simulated and real data show the  benefits of the proposed target detection and reconstruction  approaches when compared to state-of-the-art processing algorithms.
\end{abstract}

\section{Introduction}
Single-photon (SP) Light detection and ranging (LiDAR) used with their time-correlated single-photon technique is receiving significant interest as an emerging approach in numerous applications such as automotive LiDAR \cite{WallaceHalimi,rapp2020advances,lindner20093d},  remote sensing for environmental sciences \cite{wallace2013design,hakala2012full}, or imaging at long-range  \cite{tan2020deep,li2020super,mccarthy2009long,li2020single,pawlikowska2017single,li2017multi}, through obscurants   \cite{maccarone2015underwater,halimi2017object}\cite{tobin2019three} or with multiple wavelengths \cite{altmann2016efficient,halimi2019robust} \cite{halimi2021robust}. 
An SP-LiDAR system operates by illuminating the scene using laser pulses and recording the arrival times of the reflected photons with respect to their emission times using an SP detector, typically a single-photon avalanche diode (SPAD) detector. A histogram of photon counts with respect to time-of-flight is constructed for each individual SPAD pixel. By using a focal plane array composed of multiple SPAD detectors, it is possible to image the entire optical field simultaneously.  This process can be repeated using different laser wavelength illumination to obtain additional multispectral information on the observed scene. 
The resulting timing histograms enable object detection  \cite{tachella2019fast,tachella2019fast,halimi2020fast} as well as estimation of the target's depth and reflectivity profiles  \cite{belmekki2021fast,halimi2021robust}. 

However, SP-LiDAR faces several challenges to allow further deployment to real-world applications. Large data volume constitutes a major challenge for SP-LiDAR imaging due to the acquisition of millions of events per second that are typically acquired in large histogram cubes. This challenge is more evident when using multiple wavelengths to observe the scene and when
imaging through scattering media. This happens when the target return signal is attenuated and the background noise is significantly increased due to optical backscatter when imaging through atmospheric obscurants or underwater \cite{tobin2019three,maccarone2015underwater,halimi2017object, halimi2021robust}. The use of LIDAR in atmospheric obscurants means that the backscattered return will depend both on distance of the scattering event from the transceiver and the optical configuration used, typically resulting in a non-uniform background noise level in the timing histogram \cite{satat2018towards}. {A similar non-uniformity} is also observed as a result of the pile-up effect \cite{heide2018sub,pediredla2018signal} caused by the high counting rates, where the probability of a second event within a given timing period is too high, leading to an   higher background level near the beginning of the timing period.
Another limitation includes the detection of multiple-surfaces-per-pixel which usually occur  when imaging through semi-transparent materials (e.g., windows, camouflage), or in long-range depth profiling as shown in \cite{tachella2019bayesian}.

\begin{figure}[H]%
    \center%
    \includegraphics[width=0.8\textwidth]{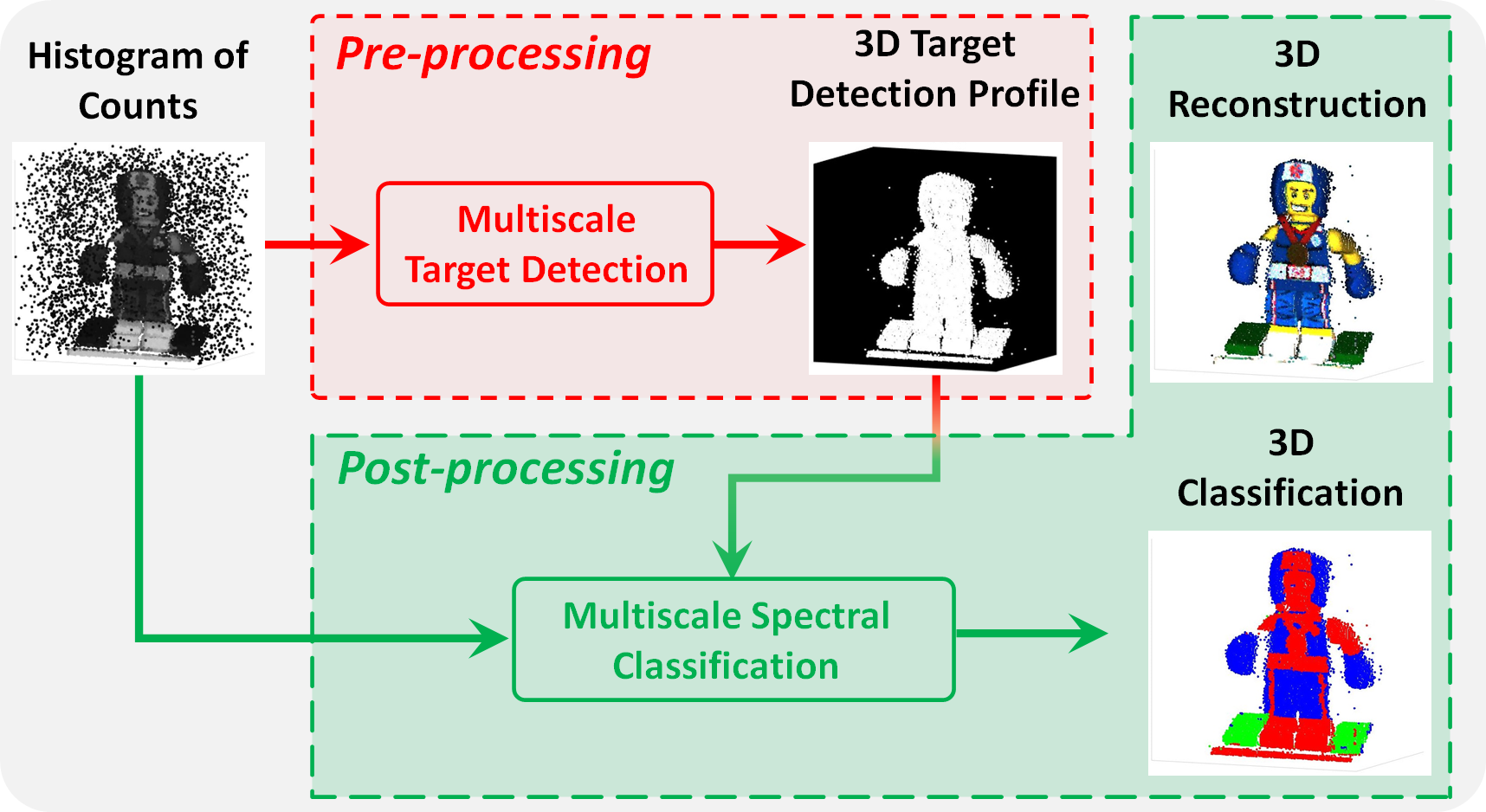}%
    \caption{Schematic representation of the main steps of the proposed framework. The top path represents the pre-processing, i.e., 3D target detection strategy. The bottom path shows the post-processing, i.e., 3D reconstruction and spectral classification strategy.}\label{fig:Pipeline}%
\end{figure}
Several approaches have been proposed to deal with these challenges. To reduce data volume, the authors of \cite{tachella2019fast,halimi2020fast} proposed a surface detection problem using a Bayesian formulation to determine if a pixel contained photons related to a surface or only spurious background counts due to ambient light or scattered photons from the observation environment (air, water, etc.). The approach in \cite{belmekki2021fast} pushed the problem further by being able to determine the material of a detected surfaces using multispectral imaging. However, these approaches assume a single peak per pixel and are operating on each pixel independently (pixel-wise), hence not taking into account spatial correlation between pixels to improve the detection.   Multiple solutions have also been proposed to improve robustness of the processing to low-light or high-background scenarios \cite{altmann2016lidar,shin2015photon,Rapp_TCI2017, halimi2016restoration}. Altmann et al. \cite{ altmann2016lidar} introduced a Bayesian approach to regularise the data fidelity term, while Shin et al. \cite{ shin2015photon}, Halimi et al. \cite{halimi2016restoration}, and Rapp and Goyal \cite{ Rapp_TCI2017} proposed  optimisation approaches as a mean to regularise the maximum likelihood solution. However, all of these approaches are not adapted to scenarios where the one-surface-per-pixel assumption is not valid. In that regard, many papers have been dedicated to tackling the multiple return problem in the context of 3D LiDAR imaging. Bayesian-based approaches \cite{hernandez2007bayesian,tachella2019bayesian} recovered the 3D profile of a target with multiple returns per-pixel. These approaches used reversible jump Markov chain Monte Carlo (RJ-MCMC) as an inference method to sample from the posterior distribution of interest and promote spatial correlation between points belonging to the same surface, showing good performance. However, the RJ-MCMC results in a large computational cost, which hinders its use for real-time imaging. 3D reconstruction from a point cloud point-of-view has also showed promising results in \cite{mallet2010marked,tachella2019real} to tackle an unknown number of objects per pixel, but these approaches do not provide uncertainty about the estimates. This shows the need for a fast and robust pre-processing algorithm, to reduce data volume and only extract useful information that can be used to perform higher level processing with quantified uncertainty.  

This paper proposes a saliency-based pre-processing algorithm to perform 3D target detection while accounting for multiscale information and the possible presence of multiple-peaks per pixel. The resulting 3D map can be used to filter the raw data to only consider useful regions hence reducing data volume and eliminating background counts. The resulting clean data can be used to perform higher level inference as a post-processing step (see Fig. \ref{fig:Pipeline}). We propose a hierarchical Bayesian model and a corresponding  estimation algorithm for spectral classification and 3D reconstruction of the detected surfaces. Comparisons with state-of-the-art algorithms show that the proposed framework  is robust and provide good results even with a high background level and in a low photon return regime. 

To summarize the main contributions of the paper are: 
\begin{itemize}
\item A new computationally efficient 3D target detection algorithm that uses multiscale information to locate unknown number of surfaces per-pixel when the background noise can be non-uniform.
\item A new hierarchical Bayesian algorithm that performs 3D spectral classification and scene reconstruction while promoting spatial  correlation between neighbouring surfaces in the 3D point-cloud.
\end{itemize}

The paper is structured as follows. The problem formulation and the proposed 3D target detection algorithm are presented in Section~\ref{sec:PropFram3DTD}. The proposed Bayesian model and estimation algorithm for 3D spectral classification and scene reconstruction are presented in section ~\ref{sec:Post-Processing}. Results on simulated and real data are analysed in Section  ~\ref{sec:Experiments}. The {conclusions} and future work are finally reported in Section \ref{sec:Conclusion}

\section{Pre-processing: 3D target detection}
\label{sec:PropFram3DTD}

\subsection{Observation model} \label{subsec:Observation model}


We consider 3D histogram cubes $\boldsymbol{Y}$ of LiDAR photon counts of dimension $N \times L \times T$, where $N$ is the number of scanned spatial positions (i.e., $N = N_{\textrm{row}} \times N_{\textrm{col}}$ pixels), $L$ is the number of spectral wavelengths and $T$ is the number of time bins. 
Akin to \cite{hernandez2007bayesian,altmann2016joint,Rapp_TCI2017}, each photon count $y_{n,l,t}$, where ${n \in \{1,...,N\}}$, $l \in \{1,...,L\}$ and $t \in \{1,...,T\}$, is assumed to follow a Poisson distribution as follows:
\begin{equation} \label{eq:4.1}
{y}_{n,l,t} | {r}_{n,1:K_s,l}, d_{n,1:K_s}, b_{n,t,l}  \sim \mathcal{P}\left\lbrace  \sum_{c=1}^{K_s}{\left[    {r}_{n,c,l}\,g_{l}(t-d_{n,c}) \right]}  +b_{n,t,l} \right\rbrace,
\end{equation}
where $\mathcal{P(.)}$ denotes a Poisson distribution, $K_s$ represents the maximum number of peaks per pixel (a pixel might have less than $K_s$ peaks), $r_{n,c,l} \geq 0$ is the spectral signature observed at the $l_{th}$ wavelength, $d_{n} \in \{1,2,...,T\}$ represents the position of an object surface at a given range from the sensor,  $b_{n,t,l} \geq 0$ is the background level associated with dark counts and ambient illumination which can be non-uniform as it depends on $t$, and $\sum_{t=1}^Tg_{l}(t)=1$ is the normalized system impulse response function (IRF), whose shape can differ between wavelength channels (assumed known from a calibration step). To deal with photon sparse regimes where pixels can have few or no photon detected, a multiscale representation was considered in several recent work by collecting photons from neighborhood pixels \cite{Rapp_TCI2017,Halimi_TCI2019}. This is equivalent to filtering the histogram cubes with a uniform kernel as in the multiscale approximation model \cite{TCI201GMS}. In this paper, we denote the filtered histograms by $\boldsymbol{Y}^{q}$  when filtered with the $q$th uniform kernel (e.g., of size $3 \times 3, 5 \times 5,$ etc). 
 

\subsection{Saliency-based target detection}
 \label{subsec:Saliency_part}



In this work, we will expand the saliency framework proposed \cite{achanta2008salient,achanta2009frequency} to high dimensional data such as LiDAR data. This method inspired several state-of-the-art algorithm that deals with RGB  \cite{mei2021camouflaged} or RGB-D data \cite{zhou2021rgb}. The approach can highlight efficiently the   salient regions of a scene while preserving objects' boundaries and discarding high frequency components due to noise, texture, or blocking artifacts.
In the context of SP Lidar, we propose to compute {the saliency matrix $\boldsymbol{S}$ of size $N  \times T$ on the observed cube of histograms, as follows:}

\begin{equation} \label{eq:4.s1}
{ \boldsymbol{S} = \left\Vert \sum^Q_{q=1} \sum_{l=1}^{L} \lambda_q (\boldsymbol{Y}^{q}_{l} 
\operatornamewithlimits{\circledast}\limits_{t}
\boldsymbol{g}_l)  -  \hat{\boldsymbol{B}}_{l} \right\Vert_{1} }
\end{equation}

where $\left\Vert . \right\Vert_{1}$ is the L1-norm, ${\bsY}^q$ is the histogram of counts of size  $N \times L \times T $ matrix, $\operatornamewithlimits{\circledast}\limits_{t}$ is the convolution operator on $t$, ${\boldsymbol{g}_l}$ is the impulse response matrix function of size $ L \times 1$, and $\hat{\boldsymbol{B}}_{l}$ is the background noise matrix of size $ N \times T$ (estimated in section  \ref{subsec:Background_estimation}). The vector $\boldsymbol{\lambda} = [\lambda_1, ..., \lambda_Q]$, whose entries sum up to one ($\sum^Q_{q=1} \lambda_q =1$), represents the contribution of each scale to the saliency of each voxel of the cube. Note that equation \eqref{eq:4.s1} is fast to compute as it only involves convolution and summation operators. 

The saliency cube in \eqref{eq:4.s1} contains voxels of background, and others originating from target peaks. To detect salient regions, we assume the saliency values of a background voxel to follow a Gamma distribution $\mathcal{G}$, and signal voxels to have positive values, as follows
\begin{eqnarray} \label{eq:gamma_distrib}
s_{n,t}  \sim & \mathcal{G} (\alpha_b, \beta_b), &  \textrm{if } (n,t) \textrm{ is a background voxel},   \nonumber \\
 &  \textrm{otherwise }   &   (n,t) \textrm{ contains a surface }.  
\end{eqnarray}

A simple hypothesis test allows us to separate signal from background voxels. This is obtained by first fitting a gamma distribution to $\bsS$ followed by a thresholding of the saliency matrix  $S >  \textrm{Thresh}(\alpha_b, \beta_b)$, where $\textrm{Thresh}(\alpha_b, \beta_b)$ depends on the gamma parameters and can be fixed based on the desired probability of false alarm. This results in a binary 3D map, denoted $\bsM$ of size $N \times T$, highlighting regions of target returns (i.e., white voxels in Fig. \ref{fig:Pipeline}) from those due to background (i.e., black voxels in Fig. \ref{fig:Pipeline}).  The saliency-based target detection algorithm is summarized in Algo. \ref{alg:Saliency_algo}.


\begin{algorithm}
\caption{Saliency-based target detection Algorithm} \label{alg:Saliency_algo}
\begin{algorithmic}[1]
       \STATE \textit{Input:} $\bsY_{l}, \bsg_{l}, l \in \left\lbrace 1,\cdots, L \right\rbrace$,   $ {Q}$ kernels 
       \STATE Compute the saliency matrix in \eqref{eq:4.s1}
       \STATE Estimate the gamma distribution in \eqref{eq:gamma_distrib}
       \STATE Threshold saliency values based on desired probability of false alarm  
       \STATE \textit{Output:} a binary cube $\bsM$ of size ($N \times T$)
\end{algorithmic}
\end{algorithm}


\subsection{Background estimation}  \label{subsec:Background_estimation}
 
Our approach requires an estimate of the background level as indicated in \eqref{eq:4.s1}.  The presence of obscurants causes the background level $b_{n,l,t}$ to be non-uniformly distributed w.r.t pixels, wavelengths, and/or time bins (depth dimension). Our background estimation strategy assumes $b_{n,l,t}$ to be smooth, which is the case if the obscurant is spatially homogeneous. This implies that the low-pass filtered histogram  $y^{Q}_{n,t,l}$, where $Q$ is the coarsest scale of the histogram, can be represented by the sum of sparse signal returns (related to the reflected target) and a 3D smooth background  $\hat{b}_{n,l,t}$. Akin to \cite{TCI201GMS}, an efficient estimation can be obtained by assuming the same temporal shape of the background for all pixels as follows
\begin{equation} \label{eq:background_3}
{\hat{b}}_{n,l,t} = \max \left( \underbar{\textit{b}}_{n,l} + \Bar{b}_{l,t} -\Bar{\Bar{b}}_{l}, 0\right) ,
\end{equation}
where {${\Bar{b}}_{l,t},  \underbar{\textit{b}}_{n,l}$ } represent the temporal and spatial shapes of the background, respectively, and  $\Bar{\Bar{b}}_{l} = \sum_t \Bar{b}_{l,t}/T $. The temporal shape can be approximated  as follows
\begin{equation} \label{eq:background_1}
{\Bar{b}}_{l,t} = median \left( y^{Q}_{\Pi_n,l,t} \right) ,
\end{equation} 
where $\Pi_n$ represents the indices of the lowest 10\% values of $y^{Q}_{n,l,t}$ to only consider background and reject signal returns, and median represents the median operator.
For a given time bin, this strategy assumes that at least 10\% of pixels only contain background without a target, which is often satisfied except when observing a perfectly lateral plane having the same depth value for all pixels. The spatial shape of the background can also be estimated as the median value over time bins, as follows: 
\begin{equation} \label{eq:background_2}
\underbar{\textit{b}}_{l,t} = median \left( y^{Q}_{n,l,:} \right).
\end{equation}








\section{Post-Processing: Bayesian 3D reconstruction and spectral classification}
\label{sec:Post-Processing}


Given the estimated 3D target map and background level, one can isolate target's peaks and 
approximate their clean histogram of counts by removing the estimated background returns (see Section \ref{subsec:Graph_estimation}). As a result, we obtain $N_s = N \times K_s$ small histograms called surfaces, where $K_s$ represents the maximum number of surfaces per pixel, i.e.,  some pixels might have less surfaces. 
These small histograms can then be used to perform higher level processing such as object recognition, classification, etc. 
In this paper, we focus on 3D reconstruction and spectral classification of the observed objects. More precisely, we assume the availability of a spectral library containing $K$ objects (defined by their $K$ spectral signatures $\bsm_k,$ for $k \in 1, ..., K$) and aim to spatially localise the objects contained in the scene. In this case, the observation model becomes
\begin{equation} \label{eq:4.1simpl}
{y}_{s,l,t} | {r}_{s,l}, d_{s}  \sim \mathcal{P}[h_s {r}_{s,l}\,g_{l}(t-d_{s}) ],
\end{equation}
where $r_{s,l}$ represents the spectral signature of the $s$th surface,  and  {$h_s$}  is a positive  coefficient allowing the model to account for different illumination conditions. Indeed, an object can be observed under different illumination conditions (e.g., shadowed regions) leading to the same spectral shape but different amplitude levels.   
It is worth-noting that the model in \eqref{eq:4.1simpl}  uses surface indices $s$ instead of pixels, where in comparison to \eqref{eq:4.1}, a surface can be associated with the $c$th peak of the $n$th pixel. Note also that  \eqref{eq:4.1simpl}  assumes the absence of the background, as it was removed in the pre-processing step.   
%
The joint likelihood, when assuming the observed surfaces, wavelengths and bins mutually independent, is then given by: 
\begin{equation} \label{eq:4.33}
{\mathit p(\bsY | \boldsymbol{R},  \bsh,  \bsd) = \prod_{l=1}^{L} \prod_{t=1}^{T} p({y_{s,l,t}} | h_{s}, r_{s,l}, d_{s})
}
\end{equation}
where $\bsh = \left(h_1,\cdots, h_S \right)$, $\bsd = \left(d_1,\cdots, d_S \right)$ and  $\boldsymbol{R}$ is a  matrix gathering $r_{s,l}, \forall s,l$.  The post-processing problem addressed in this
section consists in performing a spectral segmentation and 3D reconstruction of a target by determining the spatial class label, the spectral {response}, and the depth of each surface. This is an ill-posed inverse problem which is regularized using a Bayesian framework combining the data statistics in \eqref{eq:4.33}, with available knowledge about the unknown parameters introduced through prior distributions, as detailed below.

\subsection{Prior distributions}
\label{subsec:Prior distributions}

\textbf{Prior of $\bsR$:}
Akin to \cite{tachella2019fast,belmekki2021fast,alaa2021fast}, the classification problem intent to assign to each pixel a label associated to one of the $K$ known spectral classes. The reflectivity conjugate prior accounts for this effect by considering a mixture of $K$  gamma distributions as follows:
\begin{eqnarray} \label{eq:4.4}
\mathit P(r_{s,l}|u_s,\alpha_{k,l}^{r},\beta_{k,l}^{r},K) 
= \sum_{k=1}^{K} \delta (u_s-k) \mathcal{G}(r_{s,l};\alpha_{k,l}^{r},\beta_{k,l}^{r}) 
\end{eqnarray}

where $u_s \in \{1,...,K\}$ is a latent variable that indicates the label of the class,  $\delta(.)$ is the Dirac delta distribution centred in 0,  $\mathcal{G}(r_{s,l};\alpha_{k,l}^{r},\beta_{k,l}^{r})$ represents a gamma density with shape and scale hyperparameters $\left(\alpha_{k,l}^{r}, \beta_{k,l}^{r}\right)$. The shape hyperparameter is fixed, however the scale parameter is estimated to account for spectral variability of the signatures between pixels. Both parameters will 
encourage the mean of the estimated signatures to be around the $K$ known spectral signatures by promoting a small variance.
Unlike in \cite{tachella2019fast,belmekki2021fast,alaa2021fast} where the absence of target information is embodied in the prior distribution, the prior \eqref{eq:4.4} assumes the presence of a target as a result of the saliency-based object detection in Section \ref{sec:PropFram3DTD}.


\textbf{Prior of $\bsH$:} 
The reflectivity modulation parameter $h_s$ should be positive and is {assumed} to vary smoothly from one pixel to neighbouring ones.  These properties are promoted by assigning a hidden gamma-MRF (GMRF) \cite{rue2005gaussian} prior to $\textbf{h}$, which also ensures the tractability of the resulting posterior distribution \cite{halimi2017object,altmann2016lidar}.
More precisely, we introduce an ${N}_\textrm{row} \times N_\textrm{col} \times K_s$ auxiliary tensor with elements $w_s \in \mathbb{R}^{+} $ and define a   3D graph between $\textbf{H}$ and  ${\boldsymbol{W}}$ such that each $h_s$ is associated to 4$K_s$ elements
of ${\boldsymbol{W}}$ and vice-versa (see Fig. \ref{fig:HMB} (b)).  
The resulting joint distribution is computationally compelling as the
conditional prior distributions of $h$ and {$w$} reduce to conjugate inverse gamma ($\mathcal{IG}$) and gamma ($\mathcal{G}$) distributions as follows:
 
\begin{equation} \label{eq:4hafter}
\mathit h_{s} | W,\boldsymbol{\rho} \sim \mathcal{G}(\theta_1^h,(\theta_2^h)^{-1}),\;\; \textrm{and } \;\;   \mathit w_{s} | H,\boldsymbol{\rho} \sim \mathcal{IG}(\theta_1^{w},\theta_2^{w})
\end{equation} 
where 
$\mathit \theta_1^h = \sum^{4K_s}_{s'=1} c_{s,s'} \rho_{s',s}$,  $\; \mathit \theta_2^h = \sum^{4K_s}_{s'=1} \frac{c_{s,s'} \rho_{s',s}}{w_{s'}}
$,     $\;  \theta_1^{w} = \sum^{4K_s}_{s'=1} c_{s,s'} \rho_{s',s}$, 
$\; \theta_2^{w} = \sum^{4K_s}_{s'=1} c_{s,s'} \rho_{s',s} h_{s'}$  and $\rho_{s,s'}$ is a coupling parameter that controls the level of spatial smoothness applied by the GMRF  and $c_{s,s'}$ is a binary variable that indicates the existence or absence of a surface, where $c_{s,:} = 0$ if the $s$th surface is not detected (see Fig. \ref{fig:HMB} (b)).





\textbf{Prior of $\bsu$:}
Potts prior is assigned to the class variable $\bsu$ to promote spatial correlations between class labels.  The prior of $\bsu$ is then
obtained using the Hammersley-Clifford theorem 
\begin{equation}
f\left(\bsu\right)  =
\frac{1}{G(\gamma)} \exp\left[\gamma \sum_{s=1}^{N_s}{\sum_{s' \in \eta(s) }  \delta \left(u_{s}-u_{s'} \right)}   \right]
\label{eqt:priorLabels2}
\end{equation}
where $\gamma>0$ is the granularity coefficient, $G(\gamma)$ is a normalizing (or partition) constant and $\eta(s)$ denotes the $4 K_s$ neighbours of the $s$th surface.

\textbf{Prior of $\bsd$:}
In absence of additional knowledge, the depth parameter $d_n$ is assigned a non-informative uniform prior as follows:
 \vspace{-0.0cm}

\begin{equation} \label{eq:4.88}
\mathit p({d}_{s} = t ) =  \frac{1}{T}
 \quad , \forall{t} \in \{1,...,T\}.  
\end{equation}

Nonetheless, the depth prior can be adjusted in the event of additional information regarding the position of a target.

\textbf{Prior of $\boldsymbol{\beta}$:}
The hyper-parameter $\beta_{k,l}$  controls the mean and variance of the unknown spectral signatures $\bsR$ for each class $k$. The $\beta_{k,l}$ hyper-parameter is assigned an inverse-gamma distribution

\begin{equation} \label{eq:4.3}
\mathit \beta_{k,l} \sim \mathcal{IG}(\upsilon_{k,l},\epsilon_{k,l})
\end{equation}

where $\upsilon_{k,l}$ and $\epsilon_{k,l}$ are fixed values that are related to the mean the variance of the hyper-parameter $\beta_{k,l}$. 

\subsection{Joint Posterior distribution}

\begin{figure}[h!]
\centering \subfigure[Directed acyclic graph (DAG) of the proposed hierarchical Bayesian model. The variables inside squares are fixed, whereas the variables inside circles are estimated.]{\includegraphics[width=0.33\figwidth]{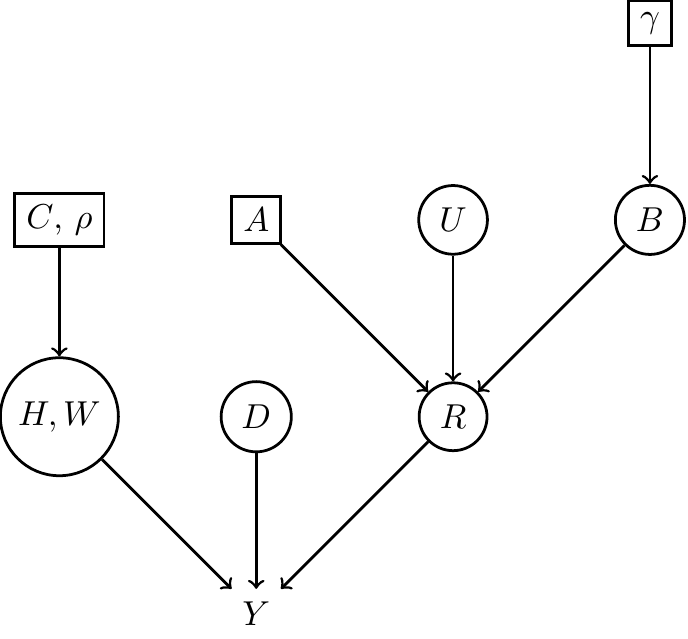}}\hspace{0.25cm}
\subfigure[Gamma-MRF neighborhood structure. Two surfaces of H and W are represented in black and red, respectively. For illustration, $w_5$ is connected to orange pixels of $h$, while $h_5$ is connected to gray pixel of $w$. Dashed orange means no detected surface for that pixel (i.e., $c=0$).]{\includegraphics[width=0.63\figwidth]{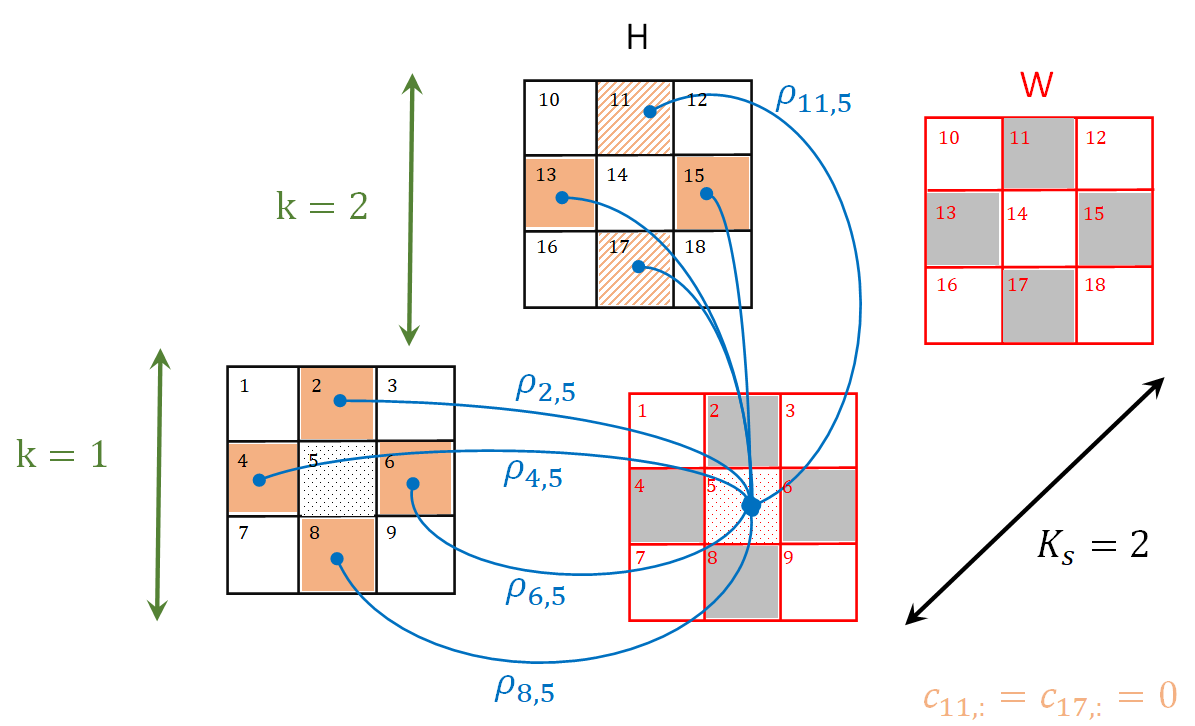}}
\caption{Connection graphs of the proposed Bayesian model and GMRF prior. } \label{fig:HMB}
\end{figure}


From the joint likelihood in (\ref{eq:4.33}) and the prior distributions considered in Section \ref{subsec:Prior distributions}, we can build the joint posterior distribution for  $\boldsymbol{\Theta}=\left(\boldsymbol{r},\boldsymbol{B},\bsd, \bsu, \bsh, \bsw \right)
$ given the 3D histograms $\boldsymbol{Y_s}$ and the hyper-parameters 
$\boldsymbol{\Theta_{H}}=\left(K, \boldsymbol{\Upsilon}, \boldsymbol{A}, \rho, \boldsymbol{C}  \right)$, where $\boldsymbol{A}$, and $\boldsymbol{\Upsilon}$ gather 
{$\{\alpha\}_{k,l}$, $\{\nu, \epsilon\}_{k,l}$}, respectively. Using Bayes rule, the joint posterior distribution can be formulated as follows:
\begin{equation}   
{\mathit p(\boldsymbol{\Theta}|\boldsymbol{Y}, \boldsymbol{\Theta_{H}}) \propto p(\boldsymbol{Y}|\boldsymbol{\Theta}) p(\boldsymbol{B}|\boldsymbol{\Upsilon},K)p(\bsH,\bsW|\boldsymbol{\rho})p(\boldsymbol{r}|\boldsymbol{A},\boldsymbol{B},K,\bsu) p(\bsu) p(\bsd). \;\; } \label{eqt:Posterior}
\end{equation}

The Directed acyclic graph (DAG) of the proposed hierarchical Bayesian model is summarized in Fig. \ref{fig:HMB}  (a).


    




\subsection{Graph estimation - Multiscale Information} \label{subsec:Graph_estimation}

This sections indicates how to obtain clean  surfaces of counts interconnected with a graph. Given the multiscale histograms $\boldsymbol{Y}^{q}_{l}$, the estimated noise $\hat{\bsB}_l$, and 3D binary detection map, one can approximate the background free signal on detected surfaces as follows  
\begin{equation} \label{eq:background_final}
 {}^{*}y_{s,l,t}^{q} = \textrm{max} \left( y^q_{s,l,t} \operatornamewithlimits{\circledast}\limits_{t} \bsg_{l} - {\hat{b}}_{s,l,t}, 0 \right) ,
\end{equation}
where $s$ represents a surface detected with the saliency based strategy in Section \ref{subsec:Saliency_part}.  
 This equation highlights the availability of multiscale information, which is often exploited in state-of-the-art algorithms to improve robustness to noise \cite{Rapp_TCI2017,halimi2019robust,tachella2019bayesian,chen2019learning,marais2017proximal,halimi2022robust,chen2020restoration}.  However, filtering the histograms of counts with a low-pass uniform filter reduces spatial resolution due to the reduction of high frequency components (edges). Different approaches for scale selection were used to pick out a scale among different available set of scales such as a median strategy or attention \cite{koo2022bayesian1,koo2022bayesian2}. Here, the goal is to select a scale $\Bar{q}$ among the $Q$ spatially downsampled version of the signal histogram of counts ${}^{*}\boldsymbol{Y}^{1:Q}$ for each surface $s$. The criterion that is applied to select a scale $\Bar{q}$ is as follows: 
\begin{equation} \label{eq:scale_sel}
\mathit {\Bar{q}_s}=\argmin\limits_{q_s \in \{2,...,Q\}} \, |d^G_s-d^{q_s}_s|
\end{equation}
where ${\Bar{q}_s}$ is the selected scale for the $s$th detected surface  and $d^G_s$ and $d^{q_s}_s$ are { given by} 
\begin{equation} \label{eq:depth_global}
\mathit {d^G_s}=\argmax\limits_{d_s \in \{1,...,T\}} \, \sum^Q_{q_s=1} \sum^L_{l=1} \sum^T_{t=1} {}^{*}\boldsymbol{y}_{s,t,l}^{q_s} g_l(t-d_s)
\end{equation}
\begin{equation} \label{eq:depth_scale}
\mathit {d^{q_s}_s}=\argmax\limits_{d_s \in \{1,...,T\}} \, \sum^L_{l=1} \sum^T_{t=1} {}^{*}\boldsymbol{y}_{s,t,l}^{q_s} g_l(t-d_s).
\end{equation}

For readability purposes, the star ${*}$ and the scale selected ${\Bar{q}_s}$ have been omitted from the denoised surface-histogram ${}^{*}y^{{\Bar{q}_s}}_{s,l,t}$ in  \eqref{eq:4.1simpl} and following equations, leading to the notation ${}^{*}y^{{\Bar{q}_s}}_{s,l,t}$ to be simply denoted ${y}_{s,l,t}$.
Note that the proposed model assumes a maximum of $K_s$ surfaces per-pixel. However, some pixels might have less surfaces in which case we introduce a binary matrix $C$ of size $N_\textrm{row} \times N_\textrm{col} \times K_s$ to indicate if a surface is detected. If a surface $s$ is not detected, $c_{s,:} = 0$ is set to $0$ and all connections to this surface will have a null weight (see Fig. \ref{fig:HMB} (b)).

\subsection{Estimation Strategy}



This paper approximates the maximum-a-posteriori parameter estimates using a  Coordinate Descent Algorithm (CDA) algorithm. The CDA aims at maximizing the posterior in \eqref{eqt:Posterior} with respect to the parameters of interest, namely $\boldsymbol{\Theta}=\left(\boldsymbol{r},\boldsymbol{B},\bsd, \bsu, \bsh, \bsw \right)$. 
Due to the large number of parameters of interest and the complexity of the joint distribution, the CDA emerges as a great candidate to estimate the parameters of interest as in  \cite{halimi2015bayesian,halimi2017denoising,halimi2017object,halimi2016hyperspectral}. Indeed, this algorithm iteratively updates the unknown parameters by maximizing their conditional distributions with respect to a single parameter while keeping the others fixed until a convergence criterion is met as detailed in Algo. \ref{alg:S_B_A_S}.


\begin{algorithm}

\def\NoNumber#1{{\def\alglinenumber##1{}\State #1}\addtocounter{ALG@line}{-1}}

\caption{Estimation Algorithm} \label{alg:S_B_A_S}
\begin{algorithmic}[1]
       \STATE \underline{Initialization}
       \STATE Initialize: $\boldsymbol{r}^{(0)}, \boldsymbol{B}^{(0)},\bsd^{(0)}, \bsu^{(0)}, \bsh^{(0)}, \bsw^{(0)}$,  
        i $\leftarrow$ 1,  conv  $\leftarrow$ 0  
        \STATE Update $\boldsymbol{d}$ with ${\boldsymbol{d}^{\bar{q}_s}}$ according to \eqref{eq:scale_sel} 
        \\
       \WHILE{conv$=0$} 
                \STATE Update $\boldsymbol{u^{(i)}}$ according to (\ref{eq:4.umax}) \\
             
                \STATE Update $\boldsymbol{r^{(i)}}$ according to \eqref{eqt:Modes_Param_top} (left)   \\
                
               \STATE Update $\boldsymbol{B^{(i)}}$ according to \eqref{eqt:Modes_Param_top} (right)\\
               
              \STATE Update $\boldsymbol{h^{(i)}}$ according to \eqref{eqt:Modes_Param_bottom} (left)\\

                \STATE Update $\boldsymbol{w^{(i)}}$ according to \eqref{eqt:Modes_Param_bottom} (right)\\ 
                
 
                
                \STATE Set conv $\leftarrow$ 1 if the convergence criteria are satisfied \\
                 
    		    \STATE i $\leftarrow$ i+1 \\
       \ENDWHILE 
\end{algorithmic}
\end{algorithm}


From  \eqref{eqt:Posterior}, we can easily demonstrate that the conditional distributions associated with the parameters of interest are

\begin{equation} \label{eq:4.r}
\mathit r_{s,l}| \boldsymbol{H}, \boldsymbol{B_{s}}, K, \boldsymbol{A}, \boldsymbol{Y_s} \sim \mathcal{G} \left[ \overline{y}_{s,l}+\alpha_{l,k},\left(h_s+\frac{1}{\beta_{k,l}}\right)^{-1} \right]
\end{equation}

\begin{equation} \label{eq:4.beta}
\mathit \beta_{s,k,l}| \boldsymbol{r}, K, \boldsymbol{A}, \boldsymbol{\Upsilon}, \boldsymbol{\rho} \sim \mathcal{IG}\left(\alpha_{l,k}+\upsilon_{l,k},r_{s,l}+\epsilon_{l,k} \right)
\end{equation}

\begin{equation} \label{eq:4.hh}
\mathit h_{s}| W, \boldsymbol{r}, \boldsymbol{\rho}, \boldsymbol{Y_s} \sim \mathcal{G} \left[\overline{\overline{y}}_{s}+L\left(\theta^h_1-1\right),\left(\sum^L_{l=1} r_{s,l}+L\theta^h_2\right)^{-1} \right]
\end{equation}

\begin{equation}  \label{eq:4.ww}
\mathit w_{s} | \mathcal{{C}},H,\boldsymbol{\rho} \sim \mathcal{IG}\left(\theta_1^{w},\theta_2^{w} \right)
\end{equation}
where $\overline{y}_{s,l}= \sum^T_{t=1} y_{s,l,t}$,  $\overline{\overline{y}}_s = \sum^L_{l=1}\overline{y}_{s,l}$
and the mode of each distribution is uniquely reached and used in the CDA updates, as follows: 

\begin{equation} \label{eq:4.umax}
\mathit u^{map}_s= \argmax\limits_{k\in \{1,...,K\}} \, p\left(u_s|\boldsymbol{H}, W, \boldsymbol{r}, \boldsymbol{B_{s}}, \boldsymbol{A}, \boldsymbol{Y_s}, \boldsymbol{\Upsilon}, \boldsymbol{\rho} \right)
\end{equation}

\begin{eqnarray} 
\mathit r^{map}_{s,l} = \frac{ \overline{y}_{s,l}+\alpha_{l,k}-1}{h_s+\frac{1}{\beta_{k,l}}},     &  \qquad \quad
\mathit \beta^{map}_{s,l,k}= \frac{r_{s,l}+\epsilon_{l,k}}{\alpha_{l,k}+\upsilon_{l,k}+1} \label{eqt:Modes_Param_top} \\
\mathit h^{map}_{s}= 
\frac{\overline{\overline{y}}_{s}+L(\theta^h_1-1)}{\sum^L_{l=1} r_{s,l}+L\theta^h_2}, 
& \qquad \quad
\mathit w^{map}_{s} =\frac{\theta_2^{w}}{\theta_1^{w}+1}.
\label{eqt:Modes_Param_bottom}
\end{eqnarray}





\subsubsection{Stopping Criteria} 

Algorithm \ref{alg:S_B_A_S} is an iterative process that requires the definition of some stopping criteria. In this paper,
we consider two criteria that would terminate the algorithm if they are satisfied. The first criterion compares the
new value of the estimated parameters $\boldsymbol{\Theta}$ to the previous ones and stops the algorithm if changes are smaller than a given user-defined threshold, that is: 

\begin{equation} \label{eq:criter}
\mathit  {\sqrt{\frac{1}{\textit{N}_s}\sum_{s=1}^{\textit{N}_s}{\Big({\pi_s^{(i+1)} - \pi_s^{(i)}}\Big)^2}}}  \leq \xi
\end{equation}

where $\textit{N}_s$ denotes the number of surfaces, $\pi_n^{(i)}$ and $\pi_n^{(i+1)}$ denote  the parameter of interest of the $s$th surface at the iterations $i$ and $i+1$, respectively.
The second criterion is based on a maximum number of iterations $I_{max}$.

\section{Experiments}
\label{sec:Experiments}

\subsection{Comparison Algorithms and Evaluation Criteria}
\label{subsec: Comparison Algorithms and Evaluation Criteria}

To highlight the robustness and benefit of the proposed algorithm, it is compared to several state-of-the-art target detection and reconstruction algorithms including:

\begin{itemize}
    \item The histogram-based TD algorithm (HTD) \cite{tachella2019fast}: is a fast target detection algorithm that operates on per-pixel histograms and assumes one surface per-pixel.
    \item The event-based TD algorithm (ETD) \cite{halimi2020fast}: 
     is a fast target detection algorithm that operates on per-pixel raw ToFs  and assumes one surface per-pixel. 
    \item The multispectral TD algorithm (MSTD) \cite{belmekki2021fast}: is a fast  per-pixel signature-based target detection algorithm that performs range estimation and multispectral classification.
    \item The multispectral 3D reconstruction algorithm (MS3D) \cite{halimi2021robust}: assumes a single-surface and designed for robust processing of multispectral LiDAR data acquired through obscurants.
    \item The RT3D algorithm \cite{tachella2019real}: assumes multiple surfaces per-pixel and is used when analysing robustness to noise and photon-sparse regime imaging on single spectral data.
    \item The MuSaPoP algorithm \cite{tachella2019bayesian}: assumes multiple surfaces per-pixel and is used when analysing multispectral LiDAR data.

\end{itemize}

Different metrics are considered to evaluate the TD and the 3D reconstruction estimates:
 
\begin{itemize}

    \item True detection $F_{\textup {true}}(\tau)$: Probability of true detection with respect to the distance $\tau$, which consider a detection as a true one if there is another point in the ground truth in the same pixel  such that $|d_{s}^{\textup {true}}$ \text{--} $d_{s'}^{\textup {est}}| \leq \tau$.

    \item False detection $F_{\textup {false}}(\tau)$: Number of estimated points that cannot be assigned to a ground truth point at a distance $\tau$.
    
    \item  Mean absolute error of intensity at distance $\tau$ denoted IAE: sum of absolute error across all detected points $S_{est}$ of intensity estimates $I^{\textup{est}}_{s,l}$  normalized with respect to the total number of ground truth points $S_{true}$, i.e., $\frac{1}{S_{true}} \sum^{S_{est}}_{s=1} \sum^L_{l=1}  |  I^{\textup{est}}_{s,l} \text{--}   I^{\textup{true}}_{s,l}|$ where $I_{s,l}=h_{s} r_{s,l}$ and $I^{\textup{true}}_{s,l}$ is the intensity of the closest ground-truth point to the estimated one. Note that if a point was falsely estimated or a ground truth point was not found, then they are considered to have resulted in an error of $\sum^{L}_{l=1} |I_{s,l}|$ .

    \item Mean absolute error depth  at distance $\tau$ denoted DAE: sum of absolute error across all detected points $S_{est}$ of depth estimates  normalized with respect to the total number of true detection $F_{\textup{true}}(\tau)$ i.e.,  $\frac{1}{F_{\textup{true}}(\tau)}  \sum^{S_{Est}}_{s=1} |  d_{s}^{\textup{true}} \text{--}   d_{s'}^{\textup{est}}|$. The ground truth and estimated points are coupled using the probability of detection $F_{\textup{true}}(\tau)$.

\end{itemize}

\subsection{Evaluation on simulated data}
\label{subsec:Evaluation_on_simulated_data}
\begin{figure}[h!]%
\center%
\includegraphics[width=0.95\textwidth]{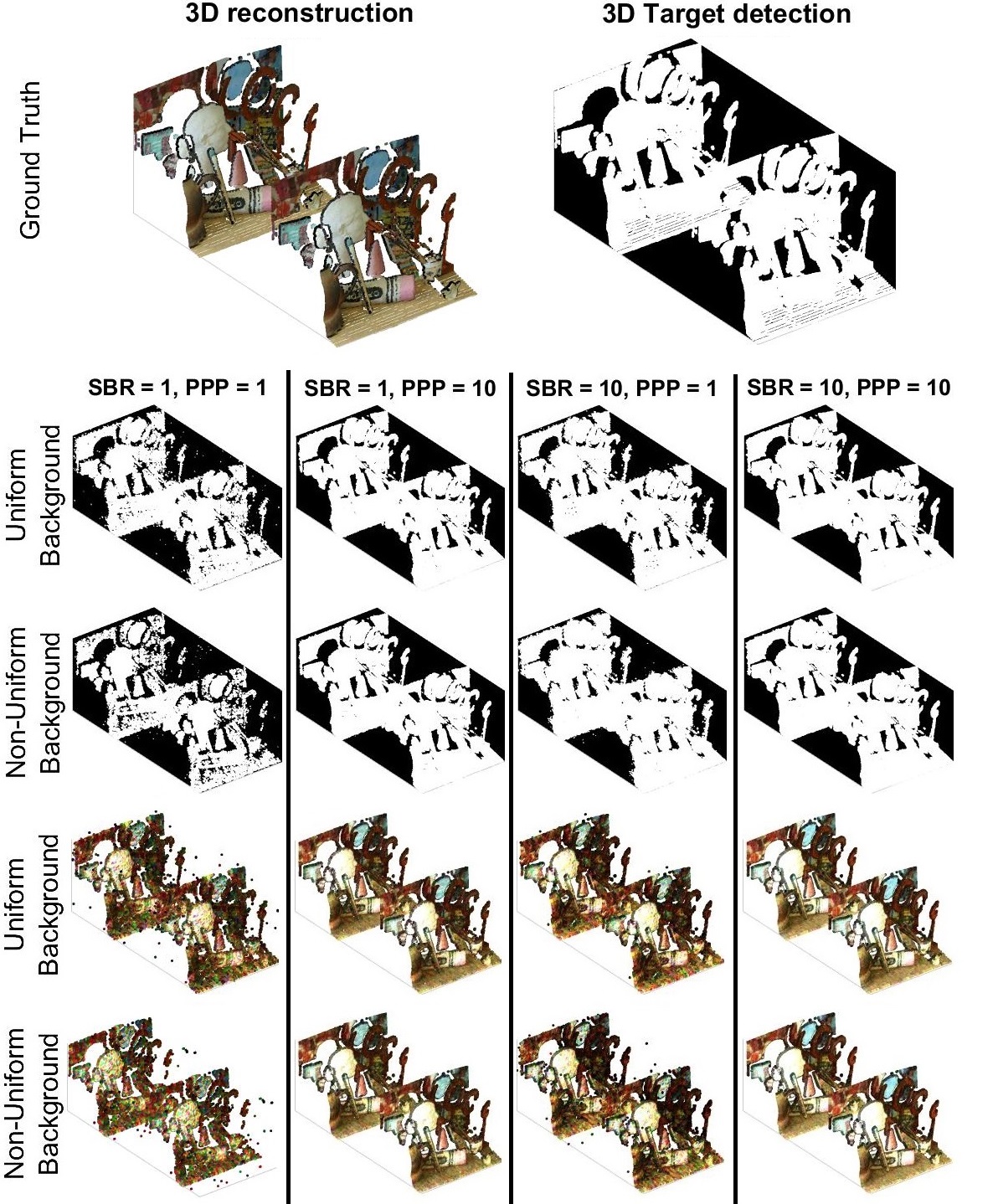}%
\caption{3D Target detection and spectral reconstruction of the proposed algorithm on the Art scene. The performance is evaluated for several levels of PPP and SBR under uniform and non-uniform background for two surfaces per-pixel. } \label{fig:3D_reconst_Art}%
\end{figure}

\begin{figure}[h!]%
\center%
\includegraphics[width=0.95\textwidth,height=6cm]{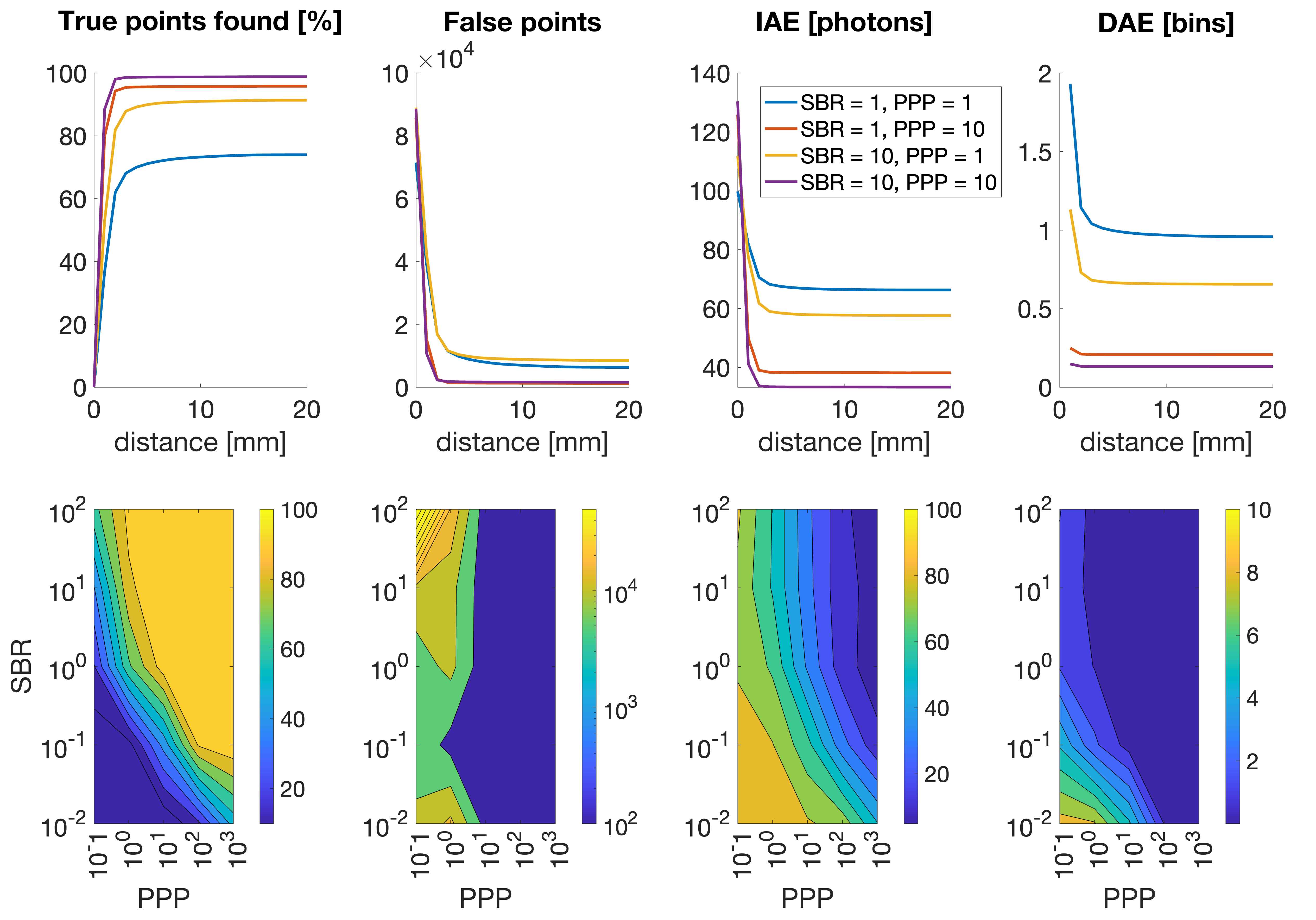}%
\caption{Quantitative results of the proposed TD and reconstruction algorithm on the Art scene. The performance is evaluated for several levels of PPP and SBR, different background shapes and for two surfaces per-pixel.}\label{fig:Quant_3D_Art}%
\end{figure}

In this section, the proposed algorithm is tested on a simulated cluttered art scene extracted from the Middlebury dataset\footnote{Available on: http://vision.middlebury.edu/stereo/data/}. This scene is commonly used as a standard  scene  for  algorithms evaluation in many  LiDAR  imaging  experiments  \cite{Rapp_TCI2017,halimi2021robust}. To simulate a scene with multiple surfaces per-pixel, two cluttered art scenes are concatenated in the depth direction. The number of pixels and bins are, respectively, $N=185\times 232$ and $T=164$. Two values of average photons per-pixel (PPP) and signal-to-background (SBR) ratio are considered and the data are corrupted by uniform and non-uniform type of background. 
The proposed algorithm has been used with $Q=4$ ($1 \times 1, 5 \times 5, 7 \times 7, 11 \times 11$), $\rho>1$, $\gamma=4$, and $\lambda_q = 1/Q, \forall q$.
Fig. \ref{fig:3D_reconst_Art} shows the qualitative results of the 3D Target profile (second and third row) and the 3D spectral reconstruction (fourth and fifth row) of the art scene.  This figure shows robust detection results in presence of the multiple surfaces per-pixel and for different levels of SBR and PPP. For uniform background, the detection results are good even at low SBR and PPP levels, but the reflectivity seems underestimated in extreme cases.   Slightly lower performance can be observed for non-uniform background at low PPP levels where the algorithm has few false detections and estimates lower reflectivity values.  Fig. \ref{fig:Quant_3D_Art} shows quantitative results for the 3D target detection 
and spectral reconstruction experiments. The first row depicts the detection performance using true, false detected points, and reconstruction results using IAE and DAE,  w.r.t distance for four values of SBR and PPP couples.  As expected, the latter metrics improves when the PPP or/and SBR increase. However, we notice that the overall performance are more sensitive to a decrease in PPP than in SBR. 
 The second row show the same performance w.r.t PPP and SBR with distance fixed to $\tau$=4mm. 
Again, the performance improves for larger PPP and/or SBR.  {However, we notice that false detection gets worse if the SBR level increases for low PPP levels.
}. 
\subsection{Evaluation on real data}
\label{subsec:Evaluation_on_real data}

This section evaluates our algorithm on different data-sets acquired with real LiDAR systems. The target detection, classification, and spectral reconstruction performance are evaluated on scenes with one and multiple surfaces per-pixel.  {In the next experimental scenarios, $Q=4$ with kernel sizes $1 \times 1, 3 \times 3, 7 \times 7, 9 \times 9$, $\rho>1$, $\gamma=4$, and $\boldsymbol{\lambda} = [0,1,0,0]$}

\subsubsection{One surface per-pixel}

\begin{figure}[h!]%
\center%
\includegraphics[width=1\textwidth]{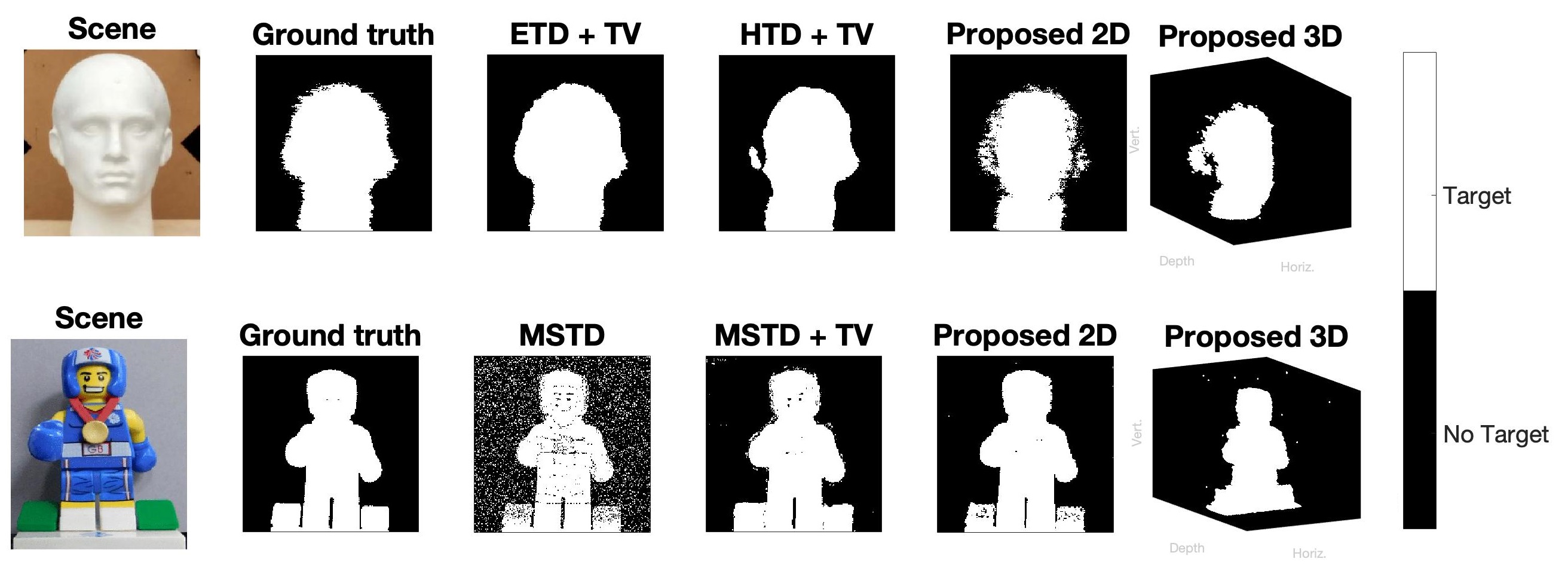}%
\caption{Comparison of the proposed TD algorithm with ETD and HTD algorithms. The scene is (top) a mannequin head with 1ms acquisition time  per-pixel (PPP $\approx$ 30), (bottom) a Lego figurine with 2ms acquisition time per-pixel (PPP $\approx$ 50).}\label{fig:Target_detection_one_surface}%
\end{figure}

\begin{figure}[h!]%
\center%
\includegraphics[width=0.95\textwidth]{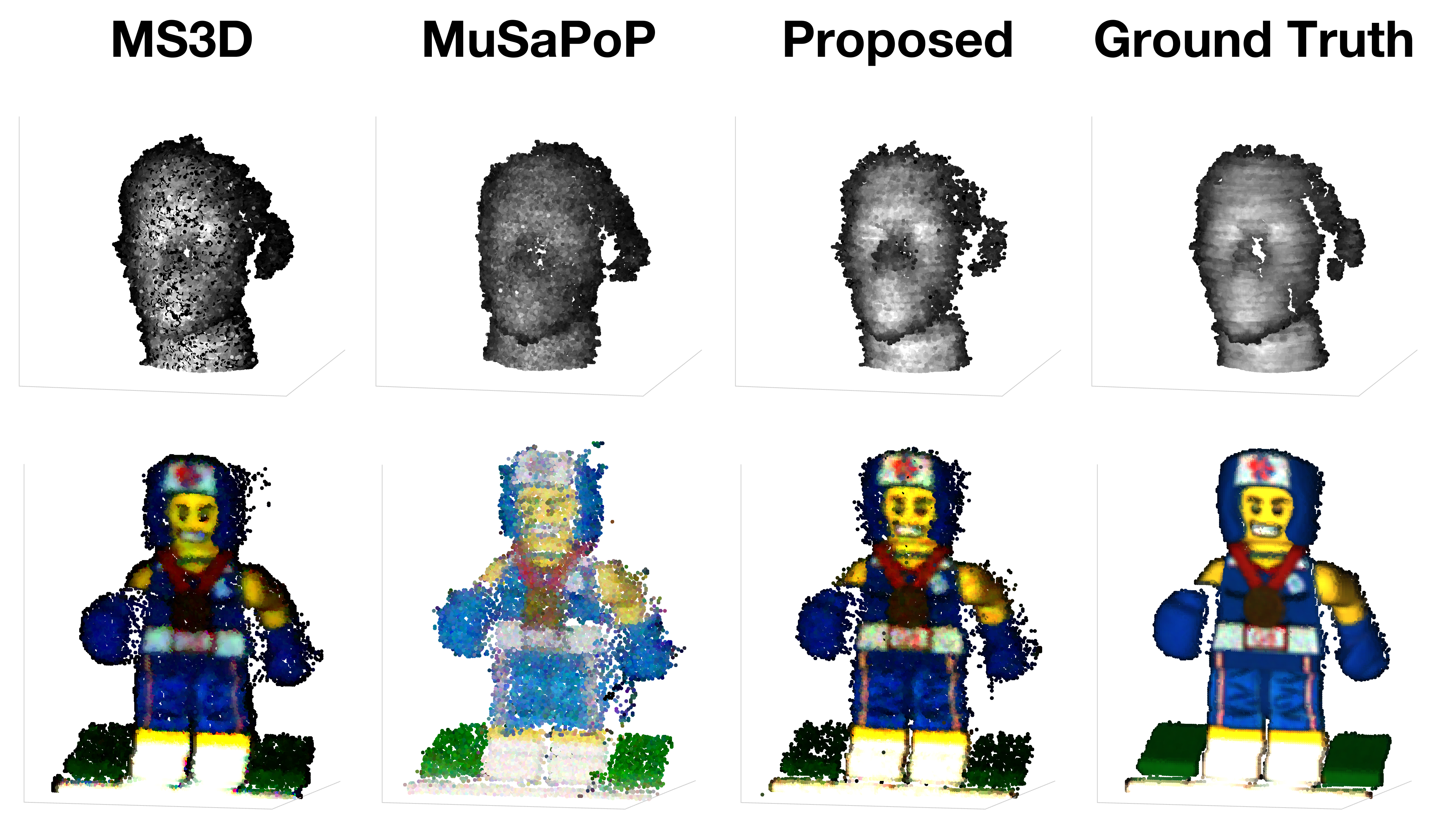}%
\caption{ {Comparison of the 3D reconstruction performance of the proposed algorithm with the MS3D, and MuSaPoP algorithms. The scene is (top) a mannequin head with 1ms acquisition time  per-pixel (PPP $\approx$ 30), (bottom) a Lego figurine with 2ms acquisition time per-pixel (PPP $\approx$ 50).}}\label{fig:3D_reconst_one_surface}%
\end{figure}

This subsection considers two real scenes that contain at most one surface per-pixel. The first scene is a  life-sized polystyrene head scanned at a stand-off distance of 325 metres during midday.  This scene consists of $N=200\times 200$ pixels, $T = 1700$ histogram bins per pixel with a timing bin size of 16ps and has an average $SBR$ of 0.25  with a 5th-95th percentile interval of (0.01, 0.6). The acquisition time per-pixel is 1 ms, which corresponds to an average PPP of 30 photons  (more details can be found in \cite{altmann2016robust} regarding this scene).  The second scene is a Lego figurine  of size 42 mm tall and 30 mm wide which was scanned at a standoff distance of 1.8 m using a TCSPC module at an acquisition time per-pixel of 160ms (40ms acquisition time per wavelength). The size of the spatial, spectral and temporal dimension of the single-photon Lego data are, respectively, $N=200 \times 200$ pixels, $L=4$ wavelengths and $ T = 1500 $ bins with a timing bin size of 2ps. The data was acquired in presence of ambient illumination leading to an SBR of $1.3$. The Lego exhibits three classes of interest ($K=3$) whose spectral signatures (related to $\boldsymbol{A}$ and $\boldsymbol{B_s}$) are extracted from pixels acquired considering a negligible background contribution and after maximum acquisition time per-pixel. The reader is referred to \cite{tobin2017comparative} for more detail regarding the Lego scene.
Fig. \ref{fig:Target_detection_one_surface} compares the estimated 3D target detection maps of the proposed algorithm with the 2D maps of the state-of-the-art TD  algorithms. Both ETD and HTD require a post-processing with a TV regularization \cite{tachella2019fast,halimi2020fast} to obtain smooth detection maps, while the proposed method provides smooth maps thanks to the use of multiscale information. Note also that ETD and HTD provide 2D maps while the proposed method detects 3D maps.  {Table \ref{tabl:AccTD} shows the accuracy of the proposed and MSTD algorithms for different acquisition times. The proposed algorithm has high accuracy performance and outperforms the MSTD algorithms post-processed with a TV regularization for an acquisition time higher than $0.4$ ms per-pixel.}
\begin{table}[H]
 \caption{ {Target detection accuracy performance of the proposed and MSTD algorithms for different acquisition times on the
Lego figurine scene.}}
 \centering
 \resizebox{.7\columnwidth}{!}{%
 \begin{tabular}{|c|c|c|c|c|c|c|}
 \hline
 Acquisition Time (ms) & 0.4 & 2 & 4 & 20 & 40 & 80 \\ \hline
 MSTD                  &     $ 73.8  $ &  $ 87.7 $  &  $ 93.3 $ & $ 97.9  $ &  $ 98.2 $  &  $ 98.2 $       \\ \hline
 MSTD + TV             &    $ \textbf{
 94.1}  $ &  $ 95.2 $  &  $ 95.7 $ & $ 98.2  $ &  $ 98.3 $  &  $ 98.2 $     \\ \hline
 Proposed              &       $ 93.3  $ &  $ \textbf{98.5} $  &  $ \textbf{98.7} $ & $ \textbf{98.6}  $ &  $ \textbf{98.9} $  &  $ \textbf{100} $      \\ \hline
 \end{tabular}
 }
 \label{tabl:AccTD}
 \end{table}
Fig. \ref{fig:3D_reconst_one_surface} shows the 3D reconstruction of the Mannequin head (first row) and the Lego figurine data (second row) for 1 ms (PPP=30) and 2ms (PPP=50) acquisition time per-pixel, respectively.  The top row of this figure shows similar  3D reconstruction  performance for all algorithms when applied to the mannequin head. For the Lego scene, we notice that the  {3D reconstruction} of the proposed algorithm and MS3D are closer to the ground truth than MuSaPoP. 
 
\begin{figure}[h!]%
\center%
\includegraphics[width=0.95\textwidth]{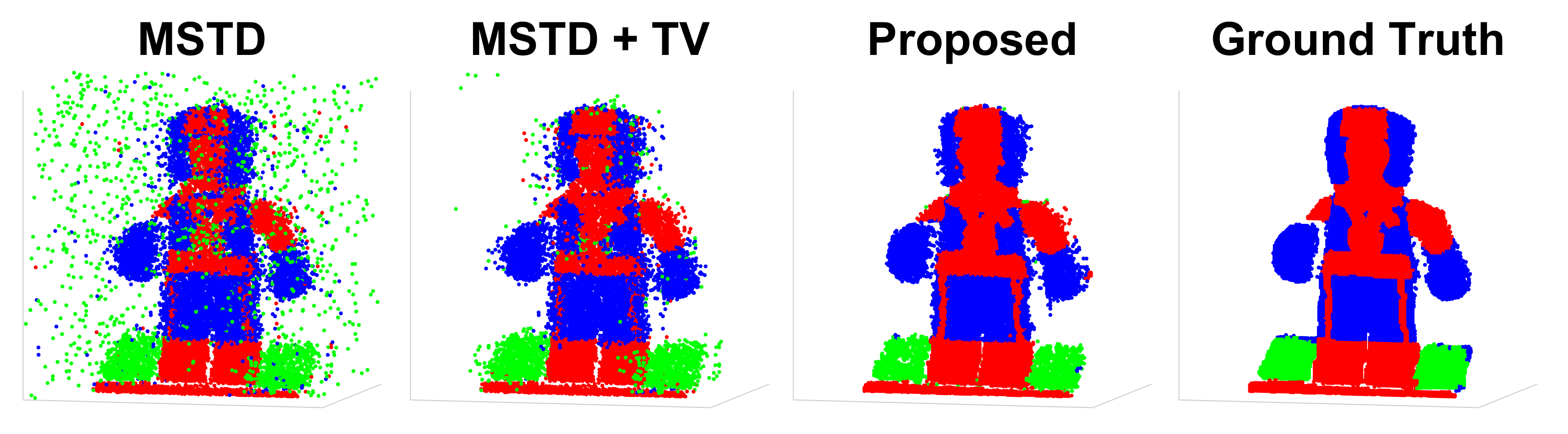}%
\caption{ {3D Spectral classification of the Lego figurine scene with 2ms acquisition time per-pixel (PPP $\approx$ 50) of the proposed and MSTD algorithms. The results of MSTD are post-processed using a TV denoiser to improve performance.  }}\label{fig:class_one_surface}%
\end{figure}

Fig. \ref{fig:class_one_surface} shows the 3D classification of the Lego scene against the MSTD algorithm. The proposed approach shows fewer false detections and better accuracy.

\subsubsection{Multiple surfaces per-pixel}

This subsection evaluates the target detection,  classification, and spectral reconstruction on real data that contain multiple surfaces per-pixel.
The first scene considered  consists of a mannequin located 4 meters behind a partially scattering object, with $N = 99 \times 99$ pixels and $T = 4000$ bins with a mean photon count per-pixel of 45. This LiDAR scene is publicly available online \footnote{https://github.com/photon-efficient-imaging/full-waveform/tree/master/fcns.
}. The second scene consists of a man standing behind camouflage at a 230 m stand-off distance from the LiDAR sensor. Different acquisition times (from 0.5 to 3.2 ms per-pixel) were used obtaining 7 to 44.6 photon per-pixel on average from which  approximately 2.1-13.3 photons correspond to background levels. The LiDAR cube has $N = 159 \times 78$ pixels and $T = 600$ histogram bins. 

\begin{figure}[h!]%
\center%
\includegraphics[width=0.95\textwidth]{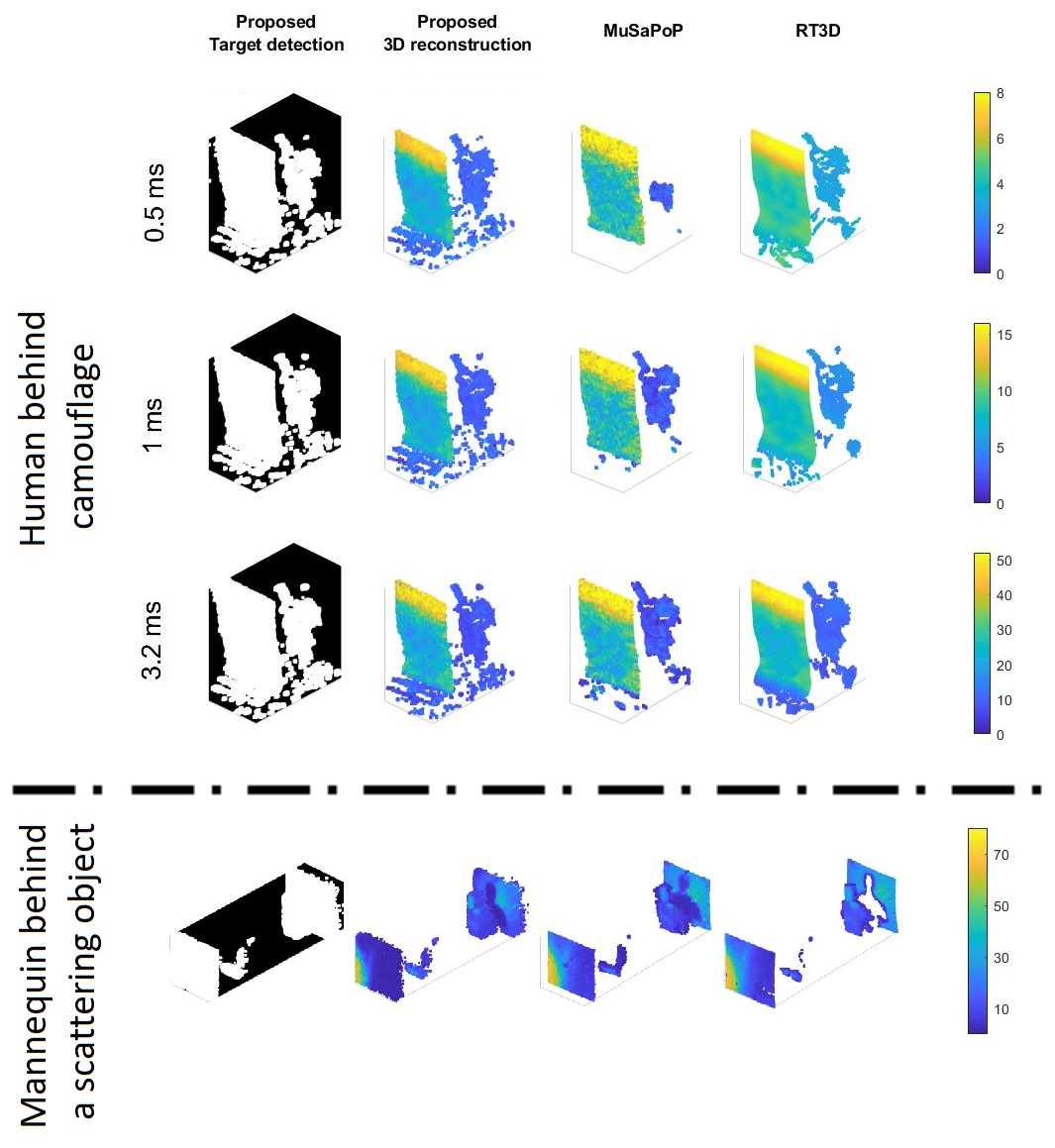}%
\caption{Comparison of the proposed 3D reconstruction algorithm with MuSaPoP, and RT3D algorithms. (1st, 2nd and 3rd rows) the scene is the man behind camouflage with different acquisition times per-pixel (0.5ms, 1ms, and 3.2ms). (4th row) the scene is a mannequin behind a scattering semi-transparent surface. From left to right, results of the 3D target detection and reconstruction of the proposed algorithm, reconstructions of MuSaPoP and RT3D. }
\label{fig:multiple_return_test}%
\end{figure}

Fig. \ref{fig:multiple_return_test} compares the proposed 3D target detection and  reconstruction with those of state-of-the-art algorithms (MuSaPoP/ManiPoP and RT3D) when considering the two real scenes.  {Different acquisition times are considered for the man behind camouflage and represented in the first three rows. The proposed algorithm shows
consistent results for these different cases and better depth estimates for small acquisition times when compared to the other state-of-the-art algorithms. 
The other scene (4th row), which depicts a mannequin man behind a scattering object, shows similar performance for all algorithms. }
Table \ref{tabl:table_TD} reports quantitative performance using the metrics defined in Section \ref{subsec: Comparison Algorithms and Evaluation Criteria} and considering as reference the proposed TD and reconstruction  estimates obtained with 3.2 ms acquisition time per-pixel.  {This table indicates best performance for the proposed approach for both true detection rates and IAE. This comes at the cost of a small increase in false alarm rate in low acquisition times.}


\begin{table}[h!]
\centering
\caption{Probabilities of true and false detection for the man behind camouflage scene for different acquisition times with   {$\tau = 1cm$}. }
\begin{tabular}{c|c|c|c|c|}
\cline{2-5}
 & Acquisition time (ms) & Proposed & MuSaPoP \cite{tachella2019bayesian} & RT3D \cite{tachella2019real}  \\ 
\hline  \multicolumn{1}{|c|}{IAE}                                   & $0.5$ & $ \textbf{ 17.72}  $ &  $ 20.45 $  &  $ 21.45 $       \\ 
\cline{2-5} \multicolumn{1}{|c|}{}                                  & $1$  & $ \textbf{14.51}  $ &  $ 16.63 $  &  $ 17.78 $    \\ 
\hline  \multicolumn{1}{|c|}{$F_{\textrm{true}}(\tau) (\%)$}          & $0.5$ &  $ \textbf{91.7}  $ &  $ 74.3 $  &  $ 81.6 $          \\ 
\cline{2-5}  \multicolumn{1}{|c|}{}                                 & $1$  & $ \textbf{94.5}  $ &  $ 87.1 $  &  $ 83.6 $     \\ 
\hline  \multicolumn{1}{|c|}{$F_{\textrm{false}}(\tau)$}              & $0.5$ & $ 243  $ &  $ \textbf{81} $  &  $ 442 $        \\ 
\cline{2-5}  \multicolumn{1}{|c|}{}                                 & $1$ &  $ \textbf{129}  $ &  $ 209 $  &  $ 254 $          \\ 
\hline
\end{tabular}
\label{tabl:table_TD}
\end{table} 



\section{Conclusions}
\label{sec:Conclusion}

This paper proposed 3D target detection and spectral classification algorithms for multispectral 3D single-photon LiDAR data. The detection algorithm exploited multiscale information to approximate the background level and detect signal peaks. This pre-processing step allowed unmixing signal information from background hence reducing data volume, and enabling higher level post-processing. In comparison to existing single-photon TD algorithms, the proposed strategy delivered smooth detection maps by considering spatial correlation between pixels, and allowed the detection of multi-peaks per pixel. We designed a hierarchical Bayesian model and estimation algorithm to perform spectral classification on the cleaned multispectral data. The results compared well with state-of-the-art 3D reconstruction algorithms on both simulated and real data. Future work will investigate the acceleration of these algorithms using parallel processing tools (e.g., GPU) to enable real-time processing.

\begin{backmatter}
\bmsection{Funding}
This work was supported by the UK Royal Academy of Engineering under the Research Fellowship Scheme (RF/201718/17128, RF/202021/20/307), and EPSRC Grants 
EP/T00097X/1, EP/S026428/1.
\end{backmatter}

\begin{backmatter}
\bmsection{Acknowledgements}
None.
\end{backmatter}

\begin{backmatter}
\bmsection{Disclosures}
The authors declare no conflicts of interest.
\end{backmatter}

\begin{backmatter}
\bmsection{Data availability}
A demo of the Matlab code will be shared after publication.
\end{backmatter}

\bibliography{biblio_all}





\end{document}